\documentstyle[12pt]{article}
\def\Journal#1#2#3#4{{#1} {\bf #2}, #3 (#4)}
\def\AP{{\em Ann. Phys.}}

\def\NPB{{\em Nucl. Phys.} B}

\def\PLB{{\em Phys. Lett.} B}

\def\PRD{{\em Phys. Rev.} D}

\newcommand{\la}{\langle}

\begin{document}

\title{THE DECONFINING PHASE TRANSITION AS AN AHARONOV-BOHM EFFECT
\footnote{To appear in the proceedings of the workshop
``Understanding Deconfinement in QCD'', Trento, Italy, March, 1999.}}

\author{JANOS POLONYI\\[0.5em]
{\small Laboratoire de Physique Th\'eorique,}
{\small Universit\'e Louis Pasteur,}\\
{\small 3, rue de l'Universit\'e, 67084 Strasbourg Cedex France,}\\
{\small and}\\
{\small Department of Atomic Physics, L. E\"otv\"os University,}\\
{\small P\'azm\'any P. S\'et\'any 1/A 1117 Budapest, Hungary}\\[0.5em]}

\maketitle
\begin{abstract}
A subjective and incomplete list of interesting
and unique features of the deconfinement phase transition
is presented. Furthermore a formal similarity of the density matrix of
the Aharonov-Bohm system and QCD is mentioned, as well.
\end{abstract}

\section{Introduction}
The first strong hints of the deconfined quarks at high temperature
appeared more than ten years ago~\cite{polsus}, the numerical
confirmation~\cite{num} followed soon. Subsequently a large
number of details has been clarified but the
driving force of the deconfinement transition, the 
confinement-deconfinement mechanism, remained elusive.
A subjective and a rather sketchy list of remarks is
presented here to indicate a unique and challenging
aspect of this phase transition. Only one detail will
be discussed in a slightly more detailed manner, 
the formal similarity between the density matrix of 
the Aharonov-Bohm (A-B) system and of QCD.

One can distinguish two confinement mechanisms~\cite{como},
a hard and a soft one. The hard mechanism is responsible
for the linear potential between static test quarks in the
absence of dynamical quarks. The soft, low energy mechanism is
which screens a test quark and was thought by V. Gribov
to be similar to the supercritical vacuum of QED. More
precisely the infrared instability of the perturbative
QCD, the source of the hard confinement mechanism, leads
to strong gluon interactions at large distances. Sufficiently 
far from a test quark the coupling constants reaches a 
large enough value to ignite the spontaneous creation
of the quark-anti quark pairs which in turn shield the
test quark charge. Most of the remarks mentioned here
refers to the hard confinement mechanism which is
more elementary and should be clarified before
embarking the study of the soft mechanism of full QCD.

\section{Unusual or unique features}
\underbar{1. Different degrees of freedom:} We find different
degrees of freedom at the two sides of the phase transition.
This happens in a number of other phase transitions,
the Mott or the localisation-delocalisation transitions
may serve as examples. Observe that the elementary degrees 
of freedom are recovered among the highly excited states
in these cases. This does not
happen in the hadronic phase. The relevance of this
obvious remark becomes clear by considering the
thermal average of an observable $A$,
$\la A\rangle=Z^{-1}\sum_n\la n|A|n\rangle e^{-{E_n/T}}$.
In order to reproduce the thermal averages we use the
color singlet asymptotic states $|n\rangle$ of the hamiltonian of the
strong interactions. How can
we recover the contributions of an isolated, deconfined
quark to the given observable? The only way out of this problem
is the modification of the Hilbert space at the phase transition.

\underbar{2. Weakly or strongly coupled phase?} 
The asymptotically free running coupling constants becomes small
around the typical energy scale $p=T$ at high energies,
$T>\Lambda_{QCD}\approx T_c$. Does that mean that
the deconfined phase is weakly coupled at high enough temperature?
The answer is known to be negative since long time~\cite{linde}.
The small parameter of the perturbation expansion at high
temperature stems from a non-perturbative quantity, the
magnetic screening mass. This can be understood by recalling that
the thermal bath breaks the Lorentz invariance.
Though the typical energy scale is pushed up at high
temperature $E\approx T$, the (off-shell) momentum scale in the
loop integrals is effected differently by the temperature and
the infrared stabilization of the long wavelength 
modes remains a difficult question. 
This is because the partition function of the 
high temperature $3+1$ dimensional 
QCD can be approximated by a $3$ dimensional (classical) 
Yang-Mills-Higgs system and the infrared sensitivity
of the partition function increases by lowering the
dimension. Thus the fate of the perturbation expansion which is
based on massless gluons depends on the screening mechanism.
The usual strategy of dealing with the IR divergences,
the separation of the scales $T$, $gT$
and $g^2T$, can not solve this problem because $g$
does not reach small enough values, $g(m_{Planck})\approx1/2$.

\underbar{3. Order parameter:} 
The order parameter related to the hard confinement mechanism 
is the trace of the heavy quark propagator continued over complex time,
\begin{equation}
\omega(\vec x,t)=\la0|\psi_\alpha\left(\vec x,t+{i\over T}\right)
\bar\psi_\alpha(\vec x,t)|0\rangle.
\end{equation}
In  the high temperature phase 
where the time extent of the Euclidean space-time is shorter
than the correlation length, $1/T<\xi\approx\Lambda^{-1}_{QCD}$,
the gluon field variables are correlated along the world line
of the heavy quark and the order parameter develops a non-vanishing
expectation value. A distinguishing feature of the deconfining
transition is that its order parameter is not a canonical variable. 
It controls the symmetry with respect the 
global center
\footnote{The center $C(G)$ of the group $G$ is a subgroup of $G$. It consists of the
elements which commute with $G$, $[C,G]=0$, e.g. 
$C(SU(N))=Z_N$.} gauge transformations performed at the initial or
the final state of a transition amplitude. It is important to keep in mind
that the center of the global gauge transformations is 
the fundamental group of the gluonic configuration space~\cite{mech},
$Z_3=\pi_1(SU(3)/Z_3)$
\footnote{Consider the gauge transformation 
$\vec A(\vec x)\to g(\vec x)(\vec\partial+\vec A(\vec x))g^\dagger(\vec x)$
acting on the anti-hermitean gauge field in the temporal gauge.
The global gauge transformations which commute with other gauge 
transformations leave $\vec A(\vec x)$ invariant.}. 
The only other known dynamical breakdown of the
fundamental group symmetry is the liquid-droplet quantum phase
transition.

\underbar{4. Finite volume effects:} 
The ratio of the
gluonic partition functions with and without a static
quark is given by the expectation value of the order
parameter, $e^{-(F_q-F)/T}=\la\omega\rangle$. Since the
spontaneous symmetry breaking does not occur in a finite system,
$\la\omega\rangle=0$ and the static quarks always appear
confined, $F_q=\infty$, in finite volume. Where does the
singular free energy density, $F_q/V=\infty$, come from?
This problem is solved by taking into account the destructive
interference between the homotopy classes in the gluonic configuration
space.

\underbar{5. Symmetry breaking by the kinetic energy:}
The spontaneous symmetry breaking mechanism is operating
at low energy where the order parameter is driven to a non-symmetrical
value due to the degenerate minima of the potential energy.
The kinetic energy might drive a spontaneous, or more precisely
dynamical symmetry breaking at high energies.
The dynamical breakdown of the center symmetry results
from such a mechanism~\cite{mech}. This can be understood 
by inspecting a quantum top, the baby version of the $SU(2)$
Yang-Mills model. The configuration space 
which consists of the $3\times3$ orthogonal matrices,
$\{R\}=SO(3)=SU(2)/Z_2$,
is doubly connected and the wave functions are single
and double valued in the integer (gluons) and the 
half-integer (quarks) spin subspaces, respectively. 
Consider now the transition amplitude 
${\cal A}(R',R)=\la R'|e^{-itH/\hbar}|R\rangle$
as the function of the
final state $R'$. Since an  orientation of the top is 
undistinguishable from its $2\pi$ rotated copy
the integer spin amplitude is doubly degenerate 
on the covering space $SU(2)$, ${\cal A}(r',r)={\cal A}(r'',r)$
where the final points $r',r''\in SU(2)$ differ in a rotation
by $2\pi$, $r'=-r''$ (center symmetry). Suppose that $r'$ is
closer to the initial point $r$ than $r''$. Then the kinetic energy tends 
to suppress the propagation to $r''$ if the time available for the 
propagation is short (high temperature or energy). The result for an infinite
top whose coordinate $r$ influences infinitely many degrees
of freedom (global gauge transformations) is that the
propagation to $r''$ is totally suppressed (center symmetry breakdown). 
The confinement can be understood as the destructive interference
in the quark propagator between the different homotopy classes.
In fact, the center symmetry of the pure gluon system yields
identical amplitudes in different homotopy classes. But a particle
in the fundamental representation of the gauge group $SU(N)$ 
propagating along the system picks up the phases $e^{2i\pi n/N}$,
$n=1,\cdots,N$
which add up to zero. The result is the absence of these particles
in the final states. We find here another 
characteristic feature of the deconfinement transition:
it corresponds to a transition amplitude rather than to the
vacuum. This is the key to find a synthesis
between the high and the low energy scattering
experiments, described in terms of the partons and the 
hadronic bound states, respectively. In other words,
as the time of a collision process is shortened the transition matrix
elements go over the ``deconfined'',
center symmetry broken phase and the elementary
constituents (partons) appear.

\underbar{6. Permanent confinement of triality~\cite{mech}:}
(i) The deconfining phase transition consists of the
dynamical breakdown of the Gauss' law and the 
modification of the Hilbert space for gluons,
\begin{equation}
H=\cases{H_0&$T<T_c,$\cr H_0\oplus H_{-1}\oplus H_1&$T>T_c$,}
\end{equation}
where the subscript stands for the triality, the center charge
\footnote{The wave functional 
$\Psi[\vec A(\vec x)]\in H_\ell$ changes by the phase factor 
$e^{2i\pi\ell n/3}$ when the global center gauge transformation
$e^{2i\pi n/3}$ is performed on $\vec A(\vec x)$. The multi-valued
nature of the wave functional is to keep track of the global center 
gauge transformations, the elements of the fundamental group of the
gluonic configuration space which are represented in a trivial manner
on the gluon field.}. 
Such a description of the phase transition is the resolution
of the puzzle mentioned in point 1.
(ii) The triality is permanently confined at any temperature.
The deconfined quark seen in the numerical simulation
is actually a composite particle containing a quark and its
vacuum polarization cloud. The latter has a multi-valued
wave functional in such a manner that the total (quark plus
gluon) wave functional is single valued. The triality charge
of the quark is screened by the unusual gluon state.
(iii) The color-magnetic monopoles relate the rotations
in the external and the color spaces. These monopoles
acquire a half-integer spin in the gluonic states with
multi-valued wave functional, a manner similar to the
generation of the spin for skyrmions. The
unusual gluonic screening cloud is the sum of
states with odd and even number of monopoles.
These components correspond to fermionic and bosonic 
exchange statistics. Thus the state of a deconfined quark is
the sum of components with bosonic (odd number of monopoles)
and fermionic (even number of monopoles) properties. The breakdown
of the center symmetry leads to the mixing of the fermi and
bose statistics for the deconfined quarks.

\underbar{7. Triality-canonical ensemble:} The transition between the
canonical and the grand-canonical ensembles requires smooth enough
dependence on the density. Due to the confinement mechanism the formal energy density
diverges for non-integer baryon numbers, or non-vanishing triality
charges (point 4.). It turns out that the triality-canonical ensemble predicts
different center domain structure at the deconfining phase transition
than the usual grand-canonical ensemble~\cite{canon}. This may happen
because the center symmetry is broken
spontaneously by the quark-anti quark see for $T<T_c$ and dynamically 
by the kinetic energy for $T>T_c$ in the canonical ensemble. This 
furthermore means that the
formal center symmetry is preserved in the presence of dynamical quarks
and the results mentioned in this talk remain valid in the triality-canonical
ensemble with dynamical quarks.

\section{Density matrix for the A-B system and for gluons}

\underbar{A-B system:}
Consider a charged particle moving on the unit circle in periodic gauge
where the wave function is periodic, $\psi(\phi+2\pi)=\psi(\phi)$.
The hamiltonian is $H=(-i\partial_\phi-\Theta/2\pi)^2/2$, where
$\Theta=2\pi A_\phi$ stands for the magnetic flux of the circle.
The eigenstates and the eigenvalues are $\psi_n(\phi)=e^{in\phi}$,
and $E_n=(n-\Theta/2\pi)^2/2$, respectively. The density matrix
is given by 
$\rho(\alpha,\beta)=Z^{-1}\sum_ne^{in(\alpha-\beta)-(n-\Theta/2\pi)^2/2T}$,
where $Z$ is the partition function, $Z=\sum_ne^{-(n-\Theta/2\pi)^2/2T}$.
Notice that the probability density $p(\phi)=\rho(\phi,\phi)$ is
real non-negative, as it should be. The periodicity of the
wave functions gives $\rho(\alpha,\alpha+2\pi)=\rho(\alpha,\alpha)$.

Let us go into an aperiodic gauge by performing the
transformation $\psi(\phi)\to e^{-i\phi\Theta/2\pi}\psi(\phi)$.
The hamiltonian is simpler, $H\to-\partial_\phi^2/2$, but has the
same spectrum as before because the wave functions are multi-valued, 
$\psi(\phi+2\pi)=e^{-i\Theta}\psi(\phi)$. In particular,
the eigenvectors are 
$\psi_n(\phi)=e^{i\phi(n-\Theta/2\pi)}$. The density matrix transforms as
$\rho(\alpha,\beta)\to e^{-i(\alpha-\beta)\Theta/2\pi}\rho(\alpha,\beta)$,
and becomes multi-valued, as well, 
$\rho(\alpha,\alpha+2\pi)=e^{i\Theta}\rho(\alpha,\alpha)$ which 
makes the construction of the probability density non-trivial.
In fact, the choice of different Riemann-sheets for the two
coordinate variables yields complex probability and partition
function. But notice that the complex factor is the same
for each contribution,
\begin{equation}
Z_{compl}=\int d\phi\rho(\phi,\phi+2\pi)=e^{i\Theta}
\int d\phi\rho(\phi,\phi),
\end{equation}
and the imaginary part of the entropy is an overall constant
which does not influence the thermalization and 
thermodynamics can be applied.

\underbar{QCD:}
A similar argument can easily be constructed for gluons 
yielding the following results: (i) The multi-valued
nature of the gluonic wave functional of a deconfined quark
is shown by the possible non-positive or complex expectation
value of the order parameter $\la\omega\rangle$, the
partition function of a quark. (ii) The density matrix
for gluons and a deconfined quark is
multi-valued as it happens for the A-B system in the aperiodic gauge.
The change of the Riemann-sheet, 
$\rho(\alpha,\alpha)\to\rho(\alpha,\alpha+2\pi)$,
corresponds to the center transformation. (iii) Thus the
complex part of the free energy and the entropy of
a deconfined quark is a simple kinematical constant which agrees for
each contribution to the partition function and does not
influence the thermalization and the applicability of the
rules of thermodynamics. (iv) The complex part of the deconfined quark
entropy may lead to observable effects in the triality-canonical
ensemble~\cite{mich} which is more realistic than the grand-canonical one.

%\section*{References}

\end{document}